\documentclass[preprint]{aastex}
\usepackage{epsfig}

\newcommand {\be} {\begin{equation}}
\newcommand {\ee} {\end{equation}}

\begin{document}

\title{Spectral Lines from Rotating Neutron Stars}

 \author{Feryal \"Ozel\altaffilmark{1} and Dimitrios Psaltis}
 \affil{Institute for Advanced Study, School of Natural Sciences, Einstein Dr., 
   Princeton, NJ 08540; \\ fozel, dpsaltis@ias.edu}
 \altaffiltext{1}{{\em Hubble} Fellow}
\slugcomment{Submitted to {\em The Astrophysical Journal Letters\/}}

\begin{abstract}

The line profiles from rotating neutron stars are affected by a number
of relativistic processes such as Doppler boosts, strong self-lensing,
frame-dragging, and the differential gravitational redshift arising
from the stellar oblateness. In this {\em Letter}, we calculate line
profiles taking into account the first two effects, which is accurate
for rotation rates less than the breakup frequency. We show that the
line profiles are not only broadened and weakened but are also
significantly asymmetric, and allow for an independent measurement of
both the mass and the radius of the neutron star. Furthermore, we
investigate the case when a fraction of the neutron star surface
contributes to the emission and find that the line profiles are
typically doubly peaked. We discuss the implications of our results
for searches for line features in the spectra of isolated neutron
stars and X-ray bursters. We finally assess the systematic
uncertainties introduced by the line asymmetry in inferring the
compactness of neutron stars from the detection of redshifted lines.

\end{abstract} 

\keywords{relativity --- stars: neutron --- X-ray: stars}

\section{Introduction}

Thermal emission from the surface of a neutron star carries signatures
of its strong gravitational field, which become apparent in
observations of both its spectral and timing properties. Such
measurements, therefore, can be used in principle to infer the masses
and radii of neutron stars.  Over the past three decades, multiple
attempts have been made to constrain the stellar equation of state
using observations of the thermal emission from bursting (Lewin, van
Paradijs, \& Taam 1995), quiescent (e.g., Rutledge et al.\ 1999), or
isolated neutron stars (e.g., Pons et al.\ 2002; Braje \& Romani
2002), the X-ray pulse profiles of rotationally powered pulsars (Page
1995; Pavlov \& Zavlin 1997), the high amplitudes of oscillations
observed during thermonuclear bursts (Nath, Strohmayer, \& Swank
2001), and the frequencies of observed quasi-periodic oscillations
(Miller, Lamb, \& Psaltis 1998).

Of all the possible methods of measuring the radius of a neutron star
or a mass-radius combination, the one that suffers the least from
systematic uncertainties and measurement errors is using the
gravitational redshift of atomic spectral lines. For a slowly
rotating, spherically symmetric neutron star, the gravitational
redshift gives directly the stellar compactness (i.e., the ratio
$R_{\rm NS}/M_{\rm NS}$). This method has received a lot of attention
recently with the launch of X-ray telescopes with high spectral
resolution (such as {\em Chandra} and {\em XMM-Newton}) and the
discovery of thermal emission from nearby, isolated neutron stars (see
Becker \& Pavlov 2002 for a review). A number of observations of
neutron stars have already been carried out, which yielded a potential
detection of broad, redshifted absorption lines from the pulsar
1E~1207.4--5209 (Sanwal et al.\ 2002; Mereghetti et al.\ 2002). Other
observations of isolated neutron stars have typically resulted in
featureless X-ray spectra (e.g., Paerels et al.\ 2001; Drake et al.\
2002; Marshall \& Schulz 2002).

Apparently, not all targets are created equal. The detection of atomic
spectral lines in X-rays requires both heavy metals to be present in
the neutron-star atmosphere and the surface layers to have high
temperatures for significant thermal emission to be generated.  These
two requirements can be met most easily in either neutron stars that
are young or in ones that are weakly magnetic and accreting steadily
from a binary companion.  Young neutron stars emit thermally the heat
released during their formation.  They may also possess heavy-element
atmospheres if significant light-element fallback did not occur during
the supernova: their short lifetimes and strong magnetic fields render
unlikely a significant accumulation of hydrogen rich material from the
interstellar medium that could suppress atomic lines. In the case of
bursters, the heavy elements in their atmospheres are continually
replenished by accretion and the thermonuclear flashes provide large
amounts of thermal energy.

Both types of neutron stars that are prime candidates for the
detection of spectral lines are fast rotators. (We do not consider
here magnetars, for which the presence of ultrastrong magnetic fields
introduces large uncertainties in calculating the rest energies of
atomic lines). The spin frequencies of known pulsars with ages
$<10^4$~yr is between $\simeq 5-65$~Hz (see, e.g., Becker \& Pavlov
2002); the inferred spin frequencies of bursters is between $\simeq
270-620$~Hz (Strohmayer 2001). These high spin frequencies introduce
several relativistic effects such as Doppler boosts, strong
self-lensing, frame-dragging and differential gravitational redshift
arising from the stellar oblateness. All of them alter the line
profiles observed at infinity.

In this {\em Letter}, we show the effects of relativistic Doppler
boosts and strong gravitational lensing on the width and asymmetry of
line profiles originating from the surfaces of rotating neutron stars.
We then investigate the systematic uncertainties introduced by these
effects in inferring the compactness of neutron stars.

\section{The Effects of Rotation on the Observed Line Profiles}

Neutron stars are the most compact stellar objects and the rotational
velocities at their surfaces can reach an appreciable fraction of the
speed of light. Therefore, rotational effects on the line profiles
originating from the neutron star surface are qualitatively different
from the case of a Newtonian slowly spinning star. In particular, the
shape of the line profiles observed at infinity is altered in four
ways.

First, relativistic Doppler boosts give rise to an asymmetry in the
spectral line profiles while broadening them. Second, strong
gravitational lensing of surface emission by the neutron star alters
the relative contribution of surface elements with different
line-of-sight velocities to the line profile. Third, frame dragging in
the rotating spacetime of the neutron star affects the photon
trajectories and thus the observable surface elements. Finally, the
stellar oblateness caused by the rapid spin introduces a difference in
the gravitational redshifts of lines that are generated at the
rotational equator and at the poles.

In this {\em Letter}, we use the numerical methods described in Muno,
\"Ozel, \& Chakrabarty (2002 and references therein) to calculate
spectral line profiles taking into account the first two of the
effects discussed above.  Investigating the latter two requires the
calculation of numerical spacetimes for specific equations of state of
neutron-star matter and will be addressed in a forthcoming paper. The
results presented here are accurate for spin frequencies smaller than
the Keplerian frequency at the surface of a neutron star, i.e., less
than a few hundred Hz.  It is important to note that for larger spin
frequencies, the ellipticity of the neutron stars induced by the
rotation can be as large as 0.3, depending on the equation of state
(Cook et al.\ 1994), yielding an additional broadening as large as
$\simeq 15$\%.

\section{Results}

In the calculations presented here, we show for clarity emission
lines; absorption lines are affected in the same way due to the linear
character of the equations. Moreover, for numerical reasons, we assume
an intrinsic fractional line width of $0.01$, which is much smaller
than the Doppler width for the cases shown here. Finally, we assume
that the observer is on the rotational equator of the star in order to
show the maximum rotational effects.

\begin{figure}[t]
\centerline{\psfig{file=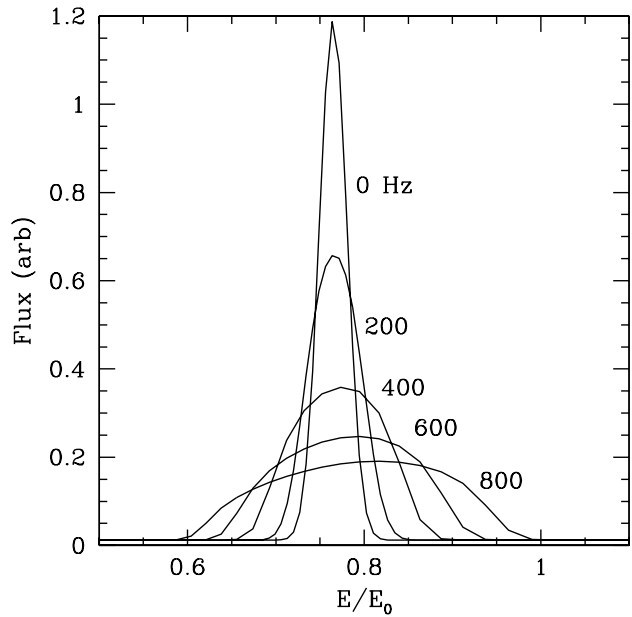,angle=0,width=10.5truecm}}
\figcaption{\footnotesize Calculated line profiles emerging from a 1.4
$M_\odot$, 10~km neutron star, for different values of its spin
frequency. Here, the entire neutron star is assumed to be emitting and
the observer is at the rotational equator. $E_0$ is the rest energy of
the emission line. }
\end{figure}

The line profiles measured at infinity coming from the surface of a
$1.4 M_\odot, 10$~km neutron star are shown in Figure~1 for different
values of the spin frequency. As the spin frequency is increased, the
lines become weaker and broader, and their peak emission shifts
towards larger energies. In this figure, the entire neutron star is 
assumed to be emitting uniformly. 

The asymmetry of the line profiles become more prominent when only a
fraction of the stellar surface contributes to the line emission.
Such a configuration is likely in the case of young neutron stars,
which may have lateral composition and temperature gradients owing to
their strong magnetic fields. Nonuniform emission is almost certainly
relevant also in the case of bursters as indicated by the observations
of large amplitude flux oscillations during the thermonuclear bursts
(Strohmayer 2001). As Figure~2 shows, in the case of non-uniform
emission, phase-integrated line profiles acquire a doubly peaked
character, with a brighter blue wing. In this figure, the emitting
region, which has an angular radius of $20^\circ$, is assumed to be at
the rotational equator. This configuration yields the largest
separation of the two peaks in the line profile because of the largest
line-of-sight velocities. Note that due to the strong gravitational
lensing, the two maxima of the doubly peaked profile do not correspond
to the line emission at rotational phases $\phi = \pm \pi/2$, where
$\phi = 0$ denotes the phase when the center of the emitting region is
aligned with the observer. As a consequence, the separation of the
peaks is not proportional to $2 \gamma \Omega R_{\rm NS} / c$, where
$\gamma$ is the Lorentz factor corresponding to the rotational
velocity and $\Omega$ is the angular velocity of the neutron star.

\begin{figure}[t]
\centerline{\psfig{file=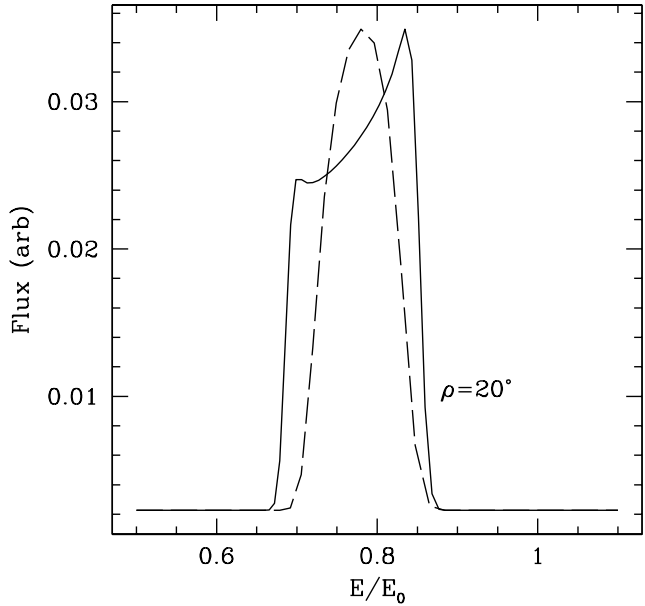,angle=0,width=10.5truecm}}
\figcaption{\footnotesize The difference in the line profiles observed
at infinity when a fraction of (solid line) or the entire neutron star
surface (dashed line) is emitting. The quantity $\rho$ denotes the
angular radius of the emitting region, which lies on the rotational
equator. The spin frequency of the neutron star is 400~Hz. Other
parameters are as in Figure~1. }
\end{figure}

The line-of-sight velocities, and thus the separation of the peaks in
the phase-averaged line profiles, are also affected by the actual size
of the neutron star, for a given stellar compactness. In Figure~3, we
show the line profiles measured at infinity that originate from two
neutron stars of the same compactness, $R/M = 2.41 G/c^2$, but of
different radii. Clearly, the larger peak separation corresponds to
the larger neutron-star radius. This effect allows in principle an
independent determination of the mass and radius (modulo emission
geometry) of a neutron star of known spin frequency, given the shape
and overall redshift of atomic spectral lines.

\begin{figure}[t]
\centerline{\psfig{file=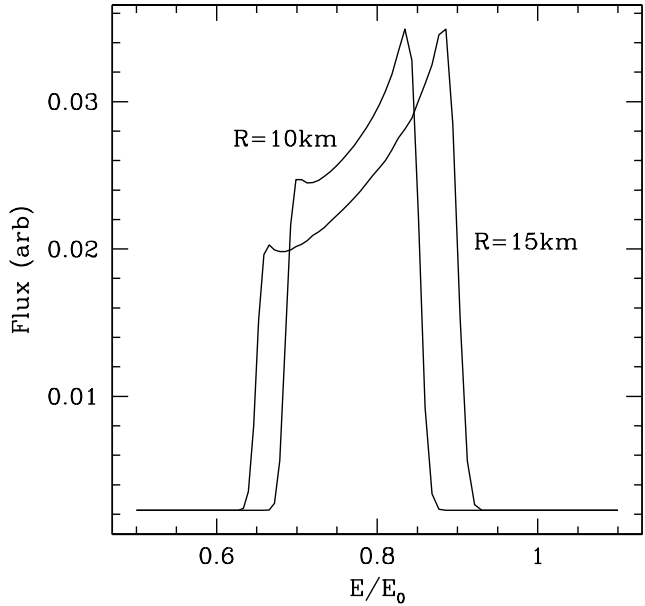,angle=0,width=10.5truecm}}
\figcaption{\footnotesize The effect of the neutron star radius on the
observed line profiles at infinity for stars of the same compactness,
$R/M = 2.41 G/c^2$.  The other parameters are the same as in
Figure~2. }
\end{figure}

\section{Discussion and Conclusions}

We studied spectral line profiles from rotating neutron stars taking
into account the effects of relativistic Doppler boosts and strong
gravitational lensing. We showed that the line profiles are broad, as
expected, and also significantly asymmetric. The asymmetry becomes
more prominent when the surface emission is non-uniform. Our results
have a number of implications for the current searches for
gravitationally redshifted line features in the spectra of neutron
stars.

First, the large widths and suppressed strengths of the rotationally
broadened lines make their detection difficult. This may be able to
account for the featureless spectra of a number of isolated neutron
stars observed with {\em Chandra} and {\em XMM-Newton}, such as
RXJ~1856--3754 (Braje \& Romani 2002; Zavlin \& Pavlov
2002). Correspondingly, if narrow line features are detected from
rapidly rotating neutron stars, e.g., bursters, they could not have
originated from the neutron star surface, unless the emission is
restricted to the rotational pole. Searches for lines in such sources
should take into account these relativistic effects.
 
\begin{figure}[t]
\centerline{\psfig{file=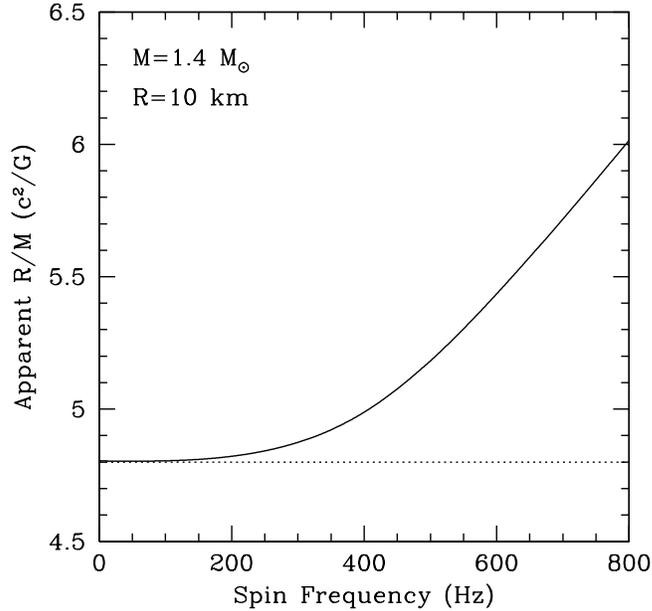,angle=0,width=10.5truecm}}
\figcaption{\footnotesize Apparent ratio $R/M$ for a neutron star,
when the peak of a spectral line is used to infer the magnitude of the
gravitational redshift; the dotted line shows the true value of
$R/M$. All the parameters are the same as in Figure~1.}
\end{figure}

Finally, the asymmetry of the line profiles introduces significant
systematic uncertainties in measuring the compactness of a neutron
star using gravitational redshifts. As an example, Figure~4 shows that
if the peak of the line is used in measuring an apparent redshift, the
resulting compactness of the neutron star will be significantly
overestimated even when the entire surface is emitting. For the
inferred spin frequencies of bursters, in the absence of realistic
models, the systematic uncertainties can be as large as $10\%$, which
are larger than the $5\%$ accuracy required to distinguish between the
different equations of state (Prakash \& Lattimer 2000).

\acknowledgements

F.\"O. acknowledges support by NASA through Hubble Fellowship grant
HF-01156 from the Space Telescope Science Institute, which is operated
by the Association of Universities for Research in Astronomy, Inc.,
under NASA contract NAS 5-26555. D.P. acknowledges the support of NSF
grant PHY-0070928.

\end{document}